\documentclass{article}
\usepackage[cp1251]{inputenc}
\usepackage[english]{babel}
\usepackage{amssymb}
\def\<{\langle}
\def\>{\rangle}

\newcommand{\Lac}{{\Lambda}_c }

\newcommand{\ksi}{{\xi}_i}

\newcommand{\tks}{{\tilde \xi}}

\newcommand{\nui}{{\nu}_i}

\newcommand{\F}{{\Phi}}

\def\thod{\theta_{12}}
\def\thdo{\theta_{21}}

\newcommand{\eti}{{\eta}_i}

\newcommand{\etiz}{{\eta}_i^0}
\newcommand{\etit}{{\eta}_i^3}

\newcommand{\Ps}{\Psi}

\newcommand{\Pszo}{\Psi_0^1}
\newcommand{\Pszi}{\Psi_0^i}

\newcommand{\vf}{\varphi }
\newcommand{\vfx}{\varphi\,(x) }

\newcommand{\vff}{\varphi_f }

\newcommand{\xz}{x_0}
\newcommand{\xo}{x_1}
\newcommand{\xd}{x_2}
\newcommand{\xt}{x_3}
\newcommand{\xch}{x_4}
\newcommand{\xit}{x_i}

\newcommand{\xitz}{x_i^0}
\newcommand{\xito}{x_i^1}
\newcommand{\xitd}{x_i^2}
\newcommand{\xitt}{x_i^3}
\newcommand{\xj}{x_j}
\newcommand{\xjz}{x_j^0}

\newcommand{\xjt}{x_j^3}

\newcommand{\wf} {W\,(x_1,  \ldots, x_n)}

\newcommand{\wfo}{W\,(x_n,  \ldots,  x_1)}
\newcommand{\wfxi}{W\,(x_1,   \ldots, x_i,  x_{i +1}, \ldots,  x_n)}
\newcommand{\wfxio}{W\,(x_1,   \ldots, x_{i + 1}, x_i, \ldots,  x_n)}

\newcommand{\wfnx}{W\,(\nu_1,   \ldots, \nu_{n - 1}, X)}

\newcommand{\wfksx}{W\,(\xi_1,   \ldots, \xi_{n - 1}, X)}
\newcommand{\vfxtsp}{\varphi \,(x_1)\,\tilde \star\, \cdots  \tilde \star\, \varphi\,(x_n) }

\newcommand{\vfitvx}{{\varphi}_i \,(t, \vec{x}) }
\newcommand{\vfotvx}{{\varphi}_1 \,(t, \vec{x}) }

\newcommand{\Pszl}{\langle \Psi_0}
\newcommand{\Pszr}{\Psi_0 \rangle}

\newcommand{\ts}{\tilde \star }

\newcommand{\dxo}{d\,x_1}

\newcommand{\dxch}{d\,x_4}

\newcommand{\cpn}{P_n}
\newcommand{\cpnz}{P_n^0}
\newcommand{\cpnt}{P_n^3}
\newcommand{\vcpn}{\vec{P_n}}
\newcommand{\ctn}{T_n}
\newcommand{\ctnmi}{T_n^-}
\newcommand{\bcvpl}{\bar{V^+}}

\newcommand{\bcvdpl}{\bar{V_2^+}}

\newcommand{\fsx}{f\,(x)}

\newcommand{\complessi}{\hbox{\rm I\hskip-5.9pt\bf C}}

\newcommand{\identity}{\hbox{\rm I\hskip-5.9pt\bf I}}
\begin{document}

\begin{center}
{ \Large{\bf{Generalized Haag's Theorem in $S \, O \, (1,1)$ and $S
\, O \, (1,3)$ Invariant Quantum Field Theory\footnote{Talk given at
the 14th International Seminar on High Energy Physics QUARKS'2006,
Repino, St.Petersburg, Russia, May 19-25, 2006 (to appear in the
Proceedings).} }}}
\end{center}

\vspace{5mm}

\begin{center}
{\large \bf M. Chaichian$^a$,  M. Mnatsakanova$^b$, A. Tureanu$^a$
 and Yu.~ Vernov$^c$}

{\it $^a$High Energy Physics Division, Department of Physical
Sciences, University of Helsinki and Helsinki Institute
of Physics, P.O. Box 64, 00014 Helsinki, Finland\\
$^b$Skobeltsyn Institute of Nuclear Physics, Moscow State
University, Moscow, Russia \\
$^c$Institute for Nuclear Research, RAS, Moscow, Russia }
\vspace{5mm}
\end{center}

\vspace{5mm}
\begin{abstract}
 {One of the most important results of the axiomatic quantum field theory - the generalized
Haag theorem - is proven in $S \, O \, (1,1)$ invariant quantum
field theory, of which an important example is noncommutative
quantum field theory. In $S \, O \, (1,3)$  invariant theory new
consequences of generalized Haag's theorem are obtained: it is
proven that the equality of four-point Wightman functions in two
theories leads to the equality of elastic scattering amplitudes and
thus total cross-sections in these theories.}
\end{abstract}

\section{Introduction}
Quantum field theory (QFT) as a mathematically rigorous and
consistent theory was formulated in the framework of the axiomatic
approach in the works of Wightman, Jost, Bogolyubov, Haag and others
(\cite{SW} - \cite{Haag}).

Within the framework of this theory, on the  basis of  most general
principles such as Poincar\'{e} invariance, local commutativity and
spectrality, a number of fundamental physical results, for example,
the CPT-theorem and the spin-statistics theorem  were proven,
analytical properties of scattering amplitudes in energy and angular
variables were established and a number of rigorous bounds on high
energy behavior of such amplitudes were obtained \cite{SW} -
\cite{BLT}, \cite{LMKh}. Haag's theorem \cite{HaagTh, FD} (see also
\cite{SW, BLT}) is one of the most novel results in the axiomatic
approach in quantum field theory (QFT). During the recent years the
different generalizations of the standard QFT have been widely
considered.

Noncommutative quantum field theory (NC QFT) is one of these
generalizations of standard QFT which has been intensively developed
during the past years (for reviews, see \cite{DN, Sz}). The idea of
such a generalization of QFT ascends to Heisenberg and it was
initially developed in Snyder’s work \cite{Snyder}. The present
development in this direction is connected with the construction of
noncommutative geometry \cite{Connes} and new physical arguments in
favour of such a generalization of QFT \cite{DFR}. Essential
interest in NC QFT is also due to the fact that in some cases it is
obtained as a low-energy limit of string theory \cite{SeWi}. The
simplest and at the same time most studied version of noncommutative
field theory is based on the following Heisenberg-like commutation
relations between coordinates:
\begin{equation} \label{cr}
[ \hat{x}_ \mu, \hat{x}_ \nu] =i \,\theta_{\mu \nu},
\end{equation}
where $\theta_{\mu\nu}$ is a constant antisymmetric matrix.

It is known that the construction of NC QFT in a general case
($\theta_{0i} \neq 0$) meets serious difficulties with unitarity and
causality \cite{GM} - \cite{CNT}. For this reason the version with
$\theta_{0i} = 0$ (space-space noncommutativity), in which there are
no such difficulties and which is a low-energy  limit of the string
theory, draws special attention. Then always there is a system of
coordinates, in which only $\thod = - \thdo \neq 0$. Thus, when
$\theta_{0i} = 0$, without loss of generality it is possible to
choose the coordinates $\xz$ and $\xt$ as commutative and the
coordinates $\xo$ and $\xd$ as noncommutative ones.

The relation (\ref{cr}) breaks the Lorentz invariance of the theory,
while the symmetry under the $SO \, (1,1) \otimes SO \, (2)$
subgroup of the Lorentz group survives \cite{AB}. Translational
invariance is still valid. Remark that when $\theta_{0i} = 0$, the
group of symmetry of the theory is $O \, (1,1) \otimes SO \, (2)$.
However, in order to obtain the results of this paper it is
sufficient to use only the weaker requirement of $SO \, (1,1)$
invariance, though we consider the case when $\theta_{0i} = 0$.
Below we shall consider the theory to be $SO \, (1,1)$ invariant
with respect to the coordinates $\xz$ and $\xt$. Besides these
classical groups of symmetry, in the paper \cite{CKNT} it was shown
that the noncommutative field theory with the commutation relation
(\ref{cr}) of the coordinates, and built according to the Weyl-Moyal
correspondence, has also a quantum symmetry, namely the twisted
Poincar\'{e} invariance.

In the works \cite{AGM} - \cite{VM05}, in which $\theta_{0i} = 0$, the
Wightman approach was formulated for NC QFT. For scalar fields the CPT theorem
and the spin-statistics theorem were proven.

In \cite{AGM} it was proposed that Wightman functions in the noncommutative
case can be written down in the standard form
\begin{equation} \label{wf} \wf = \langle \, \Psi_0, \vf (x_1)
 \ldots \vf (x_n) \, \Psi_0 \, \rangle,
\end{equation}
where $\Psi_0$ is the vacuum state. However, unlike the commutative case,
these Wightman functions are only $SO \, (1,1) \otimes SO \, (2)$ invariant.

In \cite{CMNTV} it was proposed that in the noncommutative case the
usual product of operators in the Wightman functions be replaced by
the Moyal-type product (see also \cite{Sz}). Such a product of
operators is compatible with the twisted Poincar\'{e} invariance of
the theory \cite{CPT} and also reflects the natural physical
assumption, that noncommutativity should change the product of
operators not only in coinciding points, but also  in different
ones. This follows also from another interpretation of NC QFT in
terms of a quantum shift operator \cite{CNT05}. In \cite{VM05} it
was shown that for the derivation of some axiomatic results, the
concrete type of product of operators in various points is
insignificant. It is essential only that from the appropriate
spectral condition (see formula (\ref{imp})), the analyticity of
Wightman functions with respect to the commutative variables $x^0$
and $x^3$ follows, while $x^1$ and $x^2$ do not need to be
complexified. The Wightman functions can be written down as follows
\cite{VM05}:
\begin{equation} \label{wfg}
 \wf = \Pszl, \vfxtsp \Pszr.
\end{equation}
The meaning of $\bar\star$ depends on the considered case, and in
particular
$$
\varphi \, (x) \ts \varphi \, (y) = \varphi \, (x) \varphi \, (y) \;
 \mbox{according to \cite{AGM}} \qquad \mbox{or}
$$
$$
\varphi \, (x) \ts \varphi \, (y) = \varphi \, (x) \exp{\left ({\frac{i}{2} \,
\theta_{\mu\nu} \, \frac{\overleftarrow {\partial}}{\partial x_{\mu}} \,
\frac{\overrightarrow {\partial}}{\partial y_{\nu}}} \right)} \varphi \, (y)
\; \mbox {according to \cite{CMNTV}}.
$$

In \cite{VM05} it was shown that, besides the above-mentioned theorems, in NC
QFT with ($\theta_{0i} = 0$) a number of other classical results of the
axiomatic theory, such as the equivalence of local commutativity condition and
symmetry of Wightman functions with respect to the rearrangement of the
arguments in the domain of analyticity, remain valid. In \cite{CPT} on the
basis of the twisted Poincar\'{e} invariance of the theory the Haag's theorem
was obtained \cite{HaagTh, FD} (see also \cite{SW} and references in it).

In the present work,  we consider generalized Haag's theorem (see
Theorem 4.17 in \cite{SW} or Theorem 5.4.2 in \cite{BLT}) in both $S
\, O \, (1, 3)$ and $S \, O \, ( 1, 1)$ invariant theories. An
important example of the latter case is the noncommutative theory.
In the $S \, O \, (1,3)$  invariant theory new consequences of
Haag's theorem are found, without analogues in NC QFT.

The analysis of Haag's theorem reveals essential distinctions between
commutative and noncommutative cases, more precisely between the $S \, O \,
(1,3)$ and $S \, O \, (1,1)$ invariant theories.

At the same time it is proven that the basic physical conclusion of
Haag's theorem is valid also in the $S \, O \, (1,1)$ invariant
theory. Namely, if we consider two theories, in which field
operators are connected by a unitary condition at equal times, and
one of these theories is a trivial one, that is the corresponding
$S$-matrix is equal to unity, the other is trivial as well. To
derive this result it is sufficient to assume that spectrality,
local commutativity condition and translational invariance are
fulfilled only for the transformations concerning the commutative
coordinates.

In the commutative case, the conditions which relate the field
operators by a unitary transformation at equal times (see Section
3), whose consequence is Haag's theorem, lead to the equality of
Wightman functions in the two theories up to four-point ones. In the
present paper it is shown that in the $S \, O \, (1, 1)$ invariant
theory, unlike the commutative case, only two-point Wightman
functions are equal. It is proven that from the equality of
four-point Wightman functions in the two theories, the equality of
their elastic scattering amplitudes follows and, owing to the
optical theorem, the equality of total cross sections. In derivation
of this result local coomutativity condition (LCC) is not used.

Note that actually field operators are the smoothed operators
\begin{equation}
\vff \equiv\int \,\vfx \,\fsx \, d \, x,
\end{equation}
where $\fsx$ are test functions.

Taking $\theta_{0i} = 0$, we consider a frame in which
noncommutativity does not affect the variables $\xz$ and $\xt$ and
it is possible to assume that after smearing over noncommutative
variables  the Wightman functions are standard generalized functions
(tempered distribution) with respect to the commuting coordinates,
which leads to the necessary analytical properties of Wightman
functions in these variables. It is natural to assume, as it is done
in various versions of nonlocal theory \cite{Luc} - \cite{Br}, that
with respect to the noncommutative variables Wightman functions
belong to one of Gel'fand-Shilov spaces \cite{GSh}. The concrete
choice of the Gel'fand-Shilov space is not important in the
derivation of our results.

As the consideration given below for formal operator $\varphi (x)$
coincides with the  consideration for operator $\vff$, in order not
to complicate formulas, we shall deal with the field operators as if
they were given at a point.

\section{Basic Properties of Wightman Functions }

In Wightman's approach cyclicity of the vacuum vector is assumed. In the
noncommutative case this means that any vector of the space under
consideration $J$ can be approximated with arbitrary accuracy by vectors of
the type
\begin{equation} \label{vc}
\vf (x_1) \, \ts \,\vf (x_2) \, \ts ...\ts \,\vf (x_n) \, \Psi_0.
\end{equation}
For simplicity we consider the case of a real field, however, the results are
easily extended to a complex field. For reasons given below, it is important
only that the scalar product of any two vectors $\langle \,\F, \Ps \,\rangle$
can be approximated by linear combination of Wightman functions $\wf$ with
arbitrary precision. For the results obtained below, translational invariance
only in commuting coordinates is essential, therefore we write down the
Wightman functions as:
\begin{equation} \label{ewf}
\wf = \wfksx,
\end{equation}
where $X$ designates the set of  noncommutative  variables $x_i^1, x_i^2, \; i
= 1, \ldots n$, and $\xi_i = \{\xi_i^0, \xi_i^3 \}$, where $\xi_i^0 = x_i^0 -
x_{i +1}^0, \xi_i^3 = x_i^3 - x_{i +1}^3$.

Let us formulate now the spectral condition. We assume that complete vector
system in $p$ space consists only of time-like vectors with respect to
momentum components $\cpnz$ and $\cpnt$, i.e. that
\begin{equation} \label{imp}
\cpnz \geq | \cpnt |.
\end{equation}
The condition (\ref{imp}) is conveniently written as $\cpn \in
\bcvdpl$, where $\bcvdpl$ is the set of the vectors satisfying the
condition $x^0 \geq |x^3 |$. Recall that the usual spectral
condition is $\cpn \in \bcvpl$, i.e. $\cpnz \geq | \vcpn | $. From
the condition (\ref{imp}) and the completeness of the system of
basis vectors $\Psi_{\cpn}$:
$$
\langle\Phi, \Psi\rangle = \sum_n\int d \,\cpn \langle\Phi, \Psi _{\cpn}
\rangle \langle\Psi_{\cpn}, \Psi\rangle
$$
it follows that
\begin{equation} \label{10}
\int \, d \, a \, e^{- i \, p \, a} \, \langle\Phi, U \, (a) \, \Psi\rangle =
0, \quad \mbox{if} \quad p \not \in \bar V_2^+,
\end{equation}
where $a = \{a^0, a^3 \}$ is a two-dimensional vector, $U \, (a)$ is a
translation in the plane $p^0, p^3$, and $\F$ and $\Ps$ are arbitrary vectors.
The equality (\ref{10}) is similar to the corresponding equality in the
standard case (\cite{SW}, Chap.~2.6). The direct consequence of the equality
(\ref{10}) is the spectral property of Wightman functions:
\begin{equation} \label{wfsp}
W\, (P_1..., P_{n - 1}, X) = \frac{1}{(2\pi)^{{n - 1}}} \, \int \, e^{i \, P_j
\,\xi_j} \, W \, (\xi_1..., \xi_{n - 1}, X) \, d \,\xi_1 ... d \,\xi_{n - 1} =
0,
\end{equation}
if $P_j \not \in \bar V_2^+$. The proof of the equality (\ref{wfsp})
is similar to the proof of the spectral condition in the commutative
case \cite{SW}, \cite{BLT}. Recall that in the latter case the
equality (\ref{wfsp}) is valid, if $P_j \not \in \bar V^+$. Having
written down $\wfksx$ as
\begin{equation} \label{wfspo}
W\, (\xi_1..., \xi_{n - 1}, X) = \frac{1}{(2\pi)^{{n - 1}}} \, \int \, e^{- i
\, P_j \,\xi_j} \, W \, (P_1..., P_{n - 1}, X) \, d \, P_1 ... d \, P_{n - 1},
\end{equation}
we obtain that, due to the condition (\ref{wfsp}), $\wfnx$ is analytical in
the "tube" $\; \ctnmi$:
\begin{equation} \label{tube}
\nui \in \ctnmi, \quad \mbox{if} \quad \nui = \ksi - i \,\eti, \; \eti \in
V_2^+, \; \eti = \{\etiz, \etit \}.
\end{equation}
It should be stressed that the noncommutative coordinates $\xito, \; \xitd$
are always real.

The $SO \, (1,1)$ invariance of the theory allows to expand the domain of
analyticity. This expansion is similar to the transition from tubes to
expanded tubes in the commutative case \cite{SW} - \cite{BLT}. According to
the Bargmann-Hall-Wightman theorem, $\wfnx$ is analytical in the domain $\ctn$
\begin{equation} \label{ctn}
\ctn = \cup_{\Lac} \, \Lac \,\ctnmi,
\end{equation}
where $\Lac \in S_c \, O \, (1, 1)$ is the two-dimensional analogue of the
complex Lorentz group. Just as in the commutative case, the expanded domain of
analyticity contains real points $\xit$, which are noncommutative Jost points,
satisfying the condition $\xit \sim \xj$, which means that
\begin{equation} \label{app}
{ \left (\xitz - \xjz\right)}^2 - {\left (\xitt - \xjt\right)}^2 < 0, \quad
\forall \; i, j.
\end{equation}
It should be emphasized that the noncommutative Jost points are a subset of
the set of Jost points of the commutative case, when
\begin{equation} \label{pp}
{ \left (\xit - \xj\right)}^2 < 0 \qquad \forall \; i, j.
\end{equation}

LCC has the same form as in the local theory at the Jost points (recall that
we consider a scalar field):
\begin{equation} \label{wftx}
\wf = \wfo;
\end{equation}
\begin{equation} \label{wftxi}
\wfxi = \wfxio,
\end{equation}
However, in (\ref{wftx}) and (\ref{wftxi}) $\xit$ are noncommutative
Jost points, i.e. they satisfy the condition (\ref{app}).

\section{Generalized Haag’s Theorem }

Recall the formulation of the generalized Haag’s theorem in the commutative
case (\cite{SW}, Theorem~4.17):

{\it Let $\vf_1 \, (t, \vec{x})$ and $\vf_2 \, (t, \vec{x})$ be two
irreducible sets of operators, for which the vacuum  vectors $\Psi_0^1$ and
$\Psi_0^2$ are cyclic. Further, let the corresponding Wightman functions be
analytical in the domain $\ctn$\footnote{Remark that the required analyticity
of the Wightman functions
follows only from the spectral condition and the $S \, O (1, 3)$ invariance of the theory.}. \\
Then the two-, three- and four-point Wightman functions coincide in the two
theories if there is a unitary operator $V$, such that
\begin{equation} \label{con1}
1) \quad \vf_2 \, (t, \vec{x}) = V \,\vf_1 \, (t, \vec{x}) \, V ^ *,
\end{equation}
\begin{equation} \label{con2}
2) \qquad{} \quad \Psi_0^2 = C \, V \,\Psi_0^1, \quad C \in \complessi, \quad
|C | = 1.
\end{equation}}

It should be emphasized that actually the condition  $2)$ is a
consequence of condition $1)$ with rather general assumptions (see
{\bf Statement}). In the formulation of  Haag’s theorem it is
assumed that the operators $\vfitvx$ can be smeared only on the
spatial variables. This assumption is natural if $\theta_{0i} =0 $.

Let us consider Haag’s theorem in the $S \, O (1, 1)$  invariant field theory
and show that the corresponding equality is true only for two-point Wightman
functions.

For the proof we first note that in the noncommutative case, just as in the
commutative one, from conditions  $1)$ and  $2)$ it follows that the Wightman
functions in the two theories coincide at equal times
\begin{equation} \label{32}
\langle \Psi_0^1, \vf_1 \, (t, \vec{x_1}) \, \ts \cdots \ts\, \vf_1
\, (t, \vec{x_n}) \, \Psi_0^1 \rangle = \langle \Psi_0^2, \vf_2 \,
(t, \vec{x_1}) \, \ts \cdots \ts\, \vf_2 \, (t, \vec{x_n}) \,
\Psi_0^2 \rangle.
\end{equation}

Having written down the two-point Wightman functions $W_i \, (x_1,
x_2), \; i = 1, 2$ as  $W_i \, (u_1, v_1, u_2, v_2)$, where $u_i =
\{x_i^0, x_i^3 \}, \; v_i = \{x_i^1, x_i^2 \}$ we can write for them
equality (\ref {32}) as:
\begin{equation} \label{32a}
W_1 \, (0, \xi^3, v_1, v_2) = W_2 \, (0, \xi^3, v_1, v_2),
\end{equation}
where $\xi = u_1 - u_2, \; v_1$ and $v_2$ are arbitrary vectors. Now we
notice that, due to the $S \, O (1, 1)$ invariance,
\begin{equation} \label{32b}
W_i \, (0,{\xi} ^3, v_1, v_2) = W_i \, (\tks, v_1, v_2)
\end{equation}
hence,
\begin{equation} \label{32c}
W_1 \, (\tks, v_1, v_2) = W_2 \, (\tks, v_1, v_2),
\end{equation}
where $\tks$ is any Jost point. Due to the analyticity of the Wightman
functions in the commuting variables they are completely determined by their
values at the Jost points. Thus at any $\xi$ from the equality (\ref{32c}), it
follows that
\begin{equation} \label{32d}
W_1 \, (\xi, v_1, v_2) = W_2 \, (\xi, v_1, v_2).
\end{equation}
As $v_1$ and $v_2$ are arbitrary, the formula (\ref{32d}) means the
equality of two-point Wightman functions at all values of their
arguments.

Thus, for the equality of the two-point Wightman functions in two theories
related by the conditions (\ref{con1}) and (\ref{con2}), the $S \, O (1, 1)$
invariance of the theory and corresponding spectral condition are sufficient.

It is impossible to extend this proof to three-point Wightman functions.
Indeed, let us write down $W_i \, (x_1, x_2, x_3)$ as $W_i \, (u_1, u_2, u_3,
v_1, v_2, v_3)$, where vectors $u_i$ and $v_i$ are determined  as before.
Equality (\ref{32a}) means that
\begin{equation} \label{32e}
W_1 \, (0, \xi_1^3, 0, \xi_2^3, v_1, v_2, v_3) = W_2 \, (0, \xi_1^3, 0,
\xi_2^3, v_1, v_2, v_3),
\end{equation}
$v_1, v_2, v_3$ are arbitrary. In order to have equality of the
three-point Wightman functions in the two theories from the $S \, O
(1, 1)$ invariance, the existence of transformations $\Lambda \in S
\, O (1, 1)$ connecting the points $(0, \xi_1^3)$ and $(0, \xi_2^3)$
with an open vicinity of Jost points is necessary. It would be
possible, if there existed two-dimensional vectors $\tilde{\xi}_1$
and $\tilde{\xi}_2$, $(\tilde{\xi}_i = \Lambda \, (0, \xi_i^3))$,
satisfying the inequalities:
$$
{(\tilde{\xi}_1)} ^2 < 0, \quad{(\tilde{\xi}_2)} ^2 < 0, \quad |
(\tilde{\xi}_1, \tilde{\xi}_2) | < \sqrt{{(\tilde{\xi}_1)}^2
{(\tilde{\xi}_2)}^2}.
$$
These inequalities are similar to the corresponding inequalities in the
commutative case (see equation (4.87) in \cite{SW}). However, it is easy to
check up, that the latter  from these inequalities can not be fulfilled  when
the first two are fulfilled.

Let us show now that the condition (\ref{con2}) actually is a consequence of a
condition (\ref{con1}).

{\bf  Statement} \quad{\it  Condition (\ref{con2}) is fulfilled, if the vacuum
vectors $\Pszi$ are unique, normalized, translationally invariant vectors with
respect to shifts $U_i \, (a)$ along the axis $\xt$.}

It is easy to see that the operator $U_1^{- 1} \, (a) \, V ^{- 1} \, U_2 \,
(a) \, V$ commutes with operators $\vfotvx$ and, owing to the irreducibility
of the set of these operators, it is proportional to the identity operator.
Having considered the limit $a = 0$, we see that
\begin{equation} \label{34}
U_1 ^{ - 1} \, (a) \, V ^{- 1} \, U_2 \, (a) \, V = \identity.
\end{equation}
From the equality (\ref{34}) it follows directly that if
\begin{equation} \label{35}
U_1 \, (a) \, \Pszo = \Pszo,
\end{equation}
then
\begin{equation} \label{36}
U_2 \, (a) \, V \,\Pszo = V \,\Pszo,
\end{equation}
i.e. the condition (\ref{con2}) is fulfilled. If the theory is translationally
invariant in all variables, the equality (\ref{36}) is true, if the vacuum
vector is unique, normalized, translationally invariant in the spatial
coordinates.

The most important consequence of the generalized  Haag theorem is
the following statement: if one of the two fields related by
conditions (\ref{con1}) and (\ref{con2}) is a free field, the other
is also free. In deriving this result the equality of the two-point
Wightman functions in the two theories  and LCC are used. In
\cite{CPT} it is proven that this result is valid also in the
noncommutative theory, if $\theta_{0i} = 0$.

We show below that from the equality of the four-point Wightman
functions for the fields ${\varphi}_1 \, (x)$ and ${\varphi}_2 \,
(x)$, related by the conditions (\ref{con1}) and (\ref{con2}), which
takes place in the commutative theory, an essential physical
consequence follows. Namely, for such fields the elastic scattering
amplitudes of the corresponding theories coincide, hence, due to the
optical theorem, the total cross-sections coincide as well. In
particular, if one of these fields, for example, ${\varphi}_1 \,
(x)$ is a trivial field, i.e. its corresponding $S$ matrix is equal
to unit, also the field ${\varphi}_2 \, (x)$ is trivial. In the
derivation of this result the local commutativity condition is not
used. The statement follows directly from the
Lehmann-Symanzik-Zimmermann reduction formulas \cite{LSZ}.

Let $< p_3, p_4 | p_1, p_2 >_{i}, \; i = 1,2$ be an elastic scattering
amplitudes  for the fields ${\varphi}_1 \, (x)$ and $ {\varphi}_2 \, (x)$
respectively. Owing to the reduction formulas,
$$
< p_3, p_4 | p_1, p_2 >_{i} \, \sim \, \int \,\dxo \cdots \dxch \, e ^{i \,
(-p_1 \,\xo - p_2 \,\xd + p_3 \,\xt + p_4 \,\xch)} \, \cdot
$$
\begin{equation} \label{41}
\prod_{j = 1} ^{4} \, (\square_{j} + m ^{2}) \, < 0 | T\,\varphi_{i} \, (\xo)
\, \cdots \, \varphi_{i} \, (\xch) | 0 >,
\end{equation}
where $T \,\varphi_{i} \, (\xo) \, \cdots \, \varphi_{i} \, (\xch)$ is the
chronological product of operators. From the equality
$$
W_2 \, (\xo, \ldots, \xch) = W_1 \, (\xo, \ldots, \xch)
$$
it follows that
\begin{equation} \label{42}
< p_3, p_4 | p_1, p_2 >_{1} = < p_3, p_4 | p_1, p_2 >_{2}
\end{equation}
at any $p_{i}$. Having applied this equality for the forward elastic
scattering amplitudes, we obtain that, according to the optical
theorem, the total cross-sections for the fields ${\varphi}_1 \,
(x)$ and ${\varphi}_2 \, (x)$ coincide. If now the $S$-matrix for
the field ${\varphi}_1 \, (x)$ is unity, then it is also unity for
the field ${\varphi}_2 \, (x)$. We stress that the conclusions
following from the equality of the four-point Wightman functions in
the two theories are valid in the commutative field theory.

Let us proceed now to the $S \, O (1, 1)$ symmetric theory. In this
case, using the equality of the two-point Wightman functions in the
two theories, we come to the conclusion that if  LCC is fulfilled
(\ref{wftxi}) and the current in one of the theories is equal to
zero, for example, $j_1 (x) = 0$, then $j_2 (x) = 0$ as well. Recall
that $\; j_{i} (x) = (\square + m^{2}) \, \varphi_{i} (x)$. Indeed
as $W_{1} \, (\xo, \xd) = W_{2} \, (\xo, \xd)$,
\begin{equation} \label{43}
< 0 | j_1 \, (\xo) j_1 \, (\xd) | 0 > = < 0 | j_2 \, (\xo) j_2 \, (\xd) | 0> =
0, \quad \forall \; \xo, \xd,
\end{equation}
since $j_1 \, (x) = 0$. Hence,
$$
j_2 \, (x) | 0 > = 0.
$$
In order to prove that the latter condition implies that $j_2 \, (x) = 0$ we
can use arguments similar with the ones in the standard case \cite{CMNTV}.

Let us point out that in the proof the spectral conditions  and translational
invariance only with respect to the commutating coordinates were used. If the
theory is translationally invariant with respect to all variables and the
standard spectral condition is fulfilled then if $\varphi_1 \, (x)$ is a free
field, $\varphi_2 \, (x)$ is  a free field as well \cite{CPT}.

\section{Conclusions}

It has been shown that in two $S \, O (1, 1)$ invariant theories,
whose field operators are connected by a unitary transformation at
equal times, the two-point Wightman functions coincide. This leads
to the conclusion that if one of the theories is a trivial one, then
the other is trivial as well. In the derivation of this statement
the assumptions of spectrality, local commutativity and
translational invariance were used, which are weaker than the
standard ones.

In case two $S \, O (1, 3)$ invariant theory are connected by the
same condition new consequences of Haag's theorem have been
obtained. It is proven that the equality of four-point Wightman
functions, which is fulfilled only in $S \, O (1, 3)$ invariant
theory, leads to the equality of elastic scattering amplitudes and
total cross-sections in the two theories.

\vskip0.5cm {\bf Acknowledgements} Yu.V. is supported by the grant
of the President of the Russian Federation NS-7293.2006.2
(government contract 02.445.11.7370).

\end{document}